\documentclass[pra,twocolumn,superscriptaddress]{revtex4}
\usepackage{amsbsy}
\usepackage{latexsym,epsfig,graphicx}
\usepackage{dcolumn}
\usepackage{graphicx}
\usepackage{subfigure}
\usepackage{comment}
\usepackage{color}
\usepackage{bm}
\usepackage{mathrsfs}
\usepackage{amsfonts}
\usepackage{amsmath}
\usepackage{color}
\usepackage{amssymb}
\usepackage{xspace}
\usepackage{epstopdf}
\usepackage{tabularx}
\usepackage{longtable}
\usepackage[colorlinks=true, letterpaper=true, pdfstartview=FitV, linkcolor=blue, citecolor=blue, urlcolor=blue]{hyperref}
\usepackage[normalem]{ulem}

\pdfoutput=1
\input{tcilatex}
\begin{document}

\title{Majorana corner states in an attractive quantum spin Hall insulator
with opposite in-plane Zeeman energy at two sublattice sites}
\author{Ya-Jie Wu}
\thanks{wuyajie@xatu.edu.cn}
\affiliation{School of Science, Xi'an Technological University, Xi'an 710032, China}
\author{Wei Tu}
\affiliation{School of Science, Xi'an Technological University, Xi'an 710032, China}
\author{Ning Li}
\affiliation{School of Science, Xi'an Technological University, Xi'an 710032, China}

\begin{abstract}
Higher-order topological superconductors and superfluids host
lower-dimensional Majorana corner and hinge states since novel topology
exhibitions on boundaries. While such topological nontrivial phases have
been explored extensively, more possible schemes are necessary for
engineering Majorana states. In this paper we propose Majorana corner states
could be realized in a two-dimensional attractive quantum spin-Hall
insulator with opposite in-plane Zeeman energy at two sublattice sites. The
appropriate Zeeman field leads to the opposite Dirac mass for adjacent edges
of a square sample, and naturally induce Majorana corner states. This
topological phase can be characterized by Majorana edge polarizations, and
it is robust against perturbations on random potentials as long as the edge
gap remains open. Our work provides a new possibility to realize a
second-order topological superfluid in two dimensions and engineer Majorana
corner states.
\end{abstract}

\maketitle

\section{Introduction and Motivation}

\label{S1} Higher-order topological (HOT) superconductors and superfluids
have attracted great attentions in recent years due to their novel
exhibitions of topology on the lower-dimensional boundaries including
corners and hinges \cite%
{langb2017,ZhuX2018,Yan2018,qywang2018,yxwang2018,Hsu2018,tliu2018,CZeng2019,Plekhanov2019, Bultinck2019,Peng2019,Ghorashi2019,Volpez2019,Zhang2019,Franca2019,Laubscher2019,Park2019, Niu2020,RZhang2020,Tiwari2020,Fedoseev2020,YBWu2020,YBWub2020,BRoy2020,kheir2020,Zhanga2020, MEzawa2020,Laubscher2020,Hsu2020,SBZhang2020,Ahn2020,MKheirkhah2020,XiWu2020,Laubscherb2020,Bomantara2020, Kheirkhah2020,Kheirkhah2021,Plekhanov2021,Ghosh2021,Rui2021,BFu2021,ybwu2021,ikegya2021,Jahin2021,Qin2021,ytan2021,kheirk2021,achew2021,Ghoshb2021, Yang2021,YJWu2021,BXLi2021,Zlotnikov2021,ZLi2021,XWang2021,Luo2021,YBWu2021,Tan2022}
and potential applications in topological quantum computations \cite%
{sdsar2015,pahomi2020,Song2021,Lapa2021,Pan2021,Amundsen2021}. In contrast
to conventional (first-order) topological superconductors (SCs) and
superfluids (SFs), $r$th ($r\geq 2$)-order SCs and SFs in $d$ dimensions
manifest $(d-r)$D ($d-r$-dimensional) topologically protected Majorana
boundary states. For example, second-order topological SCs and SFs in two
(three) dimensions host Majorana zero-energy modes at $0$D corners (1D
hinges).

In previous studies, a variety of HOT SCs and SFs have been proposed based
on $s_{\pm }$-wave \cite{Yan2018,qywang2018}, $p$-wave \cite{ZhuX2018}, $d$%
-wave pairings \cite{qywang2018}, and the cooperation of multi-paring orders
\cite{XZhang2019,Zhigang2019,XZhu2019}. In particular, an evidence for
helical hinge zero modes in a Fe-Based superconductor has been observed \cite%
{Gray2019}. In general, crystalline symmetries plays important roles on HOT
states \cite{Shapourian2018,Pan2019,Huang2019,Peng2020,DDVu2020,XLi2021}.
The topological classification of HOT states has been made based on spatial
symmetries \cite{khalf2018,Geier2020}. Recently, to avoid complex pairings
and complicated lattice structures, it is proposed that second-order
topological SCs and SFs could emerge in a quantum spin-Hall insulator with
the $s$-wave pairing and in-plane Zeeman field \cite{yjwu2020,YJWuJ2020}. In
these schemes, the in-plane Zeeman field gives rise to the opposite Dirac
masses on adjacent edges, and induces Majorana zero-energy states at
corners. Naturally, an interesting questions aries: are there any other
tuned parameters to implement Majorana corner states in a QSHI with s-wave
pairing besides the in-plane Zeeman field? If yes, is there a platform that
could realize this HOT phase?

In this paper, we propose that, in addition to the in-plane Zeeman field
proposed in previous proposals, the opposite in-plane Zeeman energy (OIPZE)
at two sublattice sites could also induce Majorana corner states in a
quantum spin-Hall insulator. In recent years, rapid advances in ultracold
atom systems provide a controllable platform to simulate topological quantum
states. Various parameters, such as the tunneling between sites in optical
lattices, Zeeman field, spin-orbit coupling, and the interactions between
atoms could be tuned by lasers \cite%
{Hauke2012,Aidelsburger2013,Miyake2013,Celi2014,ngold2014,Aidelsburger2015,Huang2016,ZWu2016,CChin2010,TKohler2006}%
. In particular, some topological models have been realized in this
platform, such as Haldane model \cite{Jotzu2014}, Weyl semimetals \cite%
{ZYWang2021} and etc \cite%
{Goldman2016,aeck2017,Flachner2016,Lohse2018,BSong2018}. Hence, the
ultracold atom platform could be utilized as a potential candidate to reach
the second-order topological superfluid.

In this work, we start with an attractive QSHI on a square optical lattice
with OIPZE, and derive the ground states by the mean-field theory. Through
the energy gap closing and re-opening for edge states versus OIPZE, we
deduce that the topological phase transition occurs along $y$, while not
along $x$, which implies Majorana corner states would emerge in an
appropriate parameter region. By numeric calculations, we verify the
second-order topological SF phase and characterize it by the Majorana edge
polarizations. To more thoroughly understand the emergence of Majorana
corner states, we investigate this system through the low-energy edge theory.
It is found that the cooperation of the OIPZE and s-wave pairing order could
also leads to Dirac mass at adjacent edges possessing opposite signs.
However, contrary to the in-plane Zeeman field in Refs. \cite%
{yjwu2020,YJWuJ2020}, OIPZE determines the sign of Dirac mass along $y$
while not along $x$.

The remainder of this paper is organized as follows. In Sec. \ref{S2}, we
introduce an attractive quantum spin-Hall insulator (QSH) on a square
lattice. In Sec. \ref{S3}, we study its ground states by mean-field theory
and present its global phase diagram. In Sec. \ref{S4}, we utilize Majorana
edge polarizations to characterize the second-order topological superfluid
in the phase diagram. In Sec. \ref{S5}, we explore the origin of the
emergence of Majorana corner states by the effective low-energy edge theory.
Finally, we draw discussions and conclusions in Sec. \ref{S6}.

\section{Correlated quantum spin-Hall insulator in square lattices}

\label{S2} 
\begin{figure}[tbp]
\centering\includegraphics[width=0.47\textwidth]{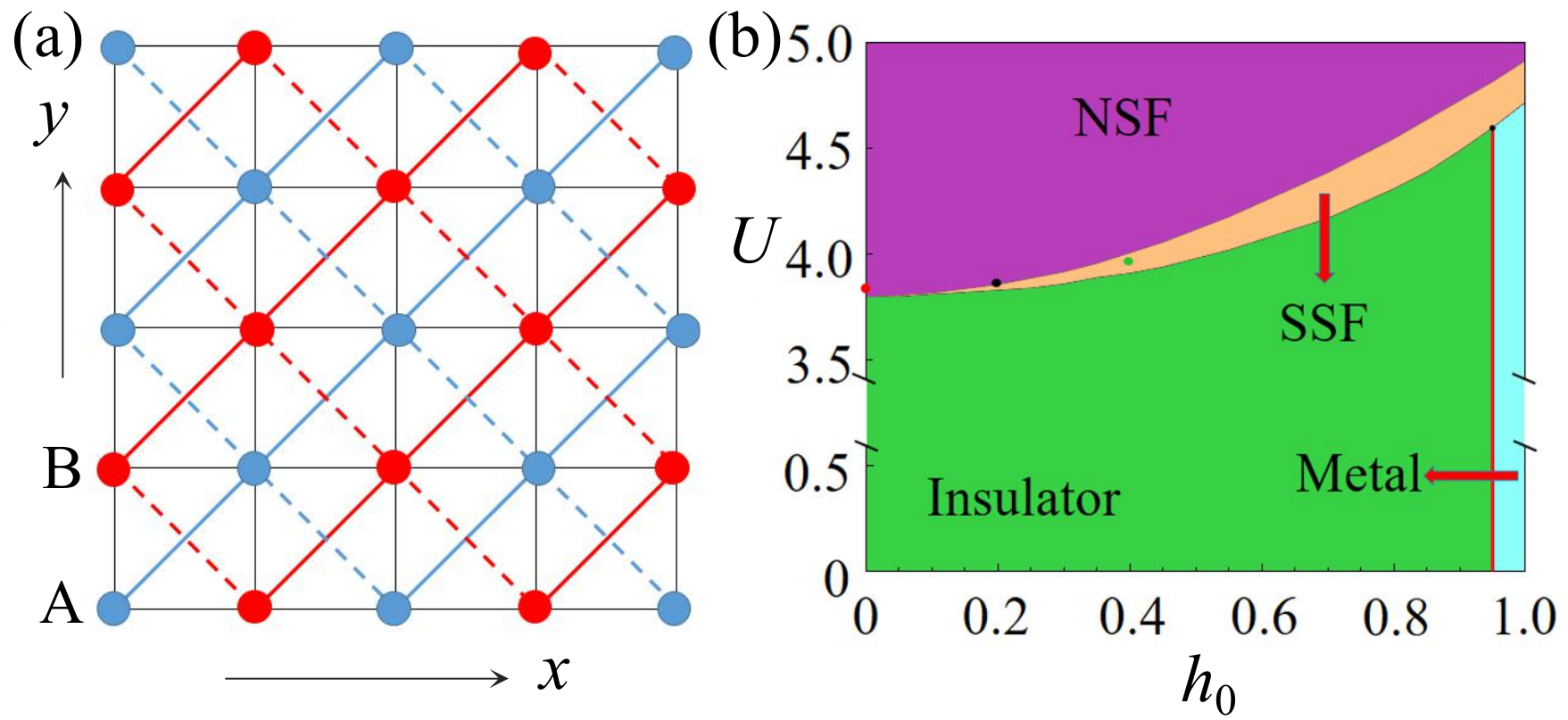}
\caption{(a) Illustration of quantum spin Hall insulator on a square
lattice. The fermion acquires a phase $\protect\pi$ hopping along dashed
lines while a phase $0$ hopping along solid lines. $\text{A}$ and $\text{B}$
denote two sublattice sites. (b) Global phase diagram. NSF denotes a normal
superfluid, and SSF indicates a second-order topological superfluid.
Parameters are set to be $t_{1}=1$, $t_{2}=0.25$, $m_{0}=0.05$ and $h_{0}=%
\protect\mu_{0}=0$.}
\label{phases}
\end{figure}

We consider a two-component Fermi gas loaded into a square
optical lattice as sketched in Fig. \ref{phases} (a). Its physics is
described by the following effective Hamiltonian as
\begin{eqnarray}
H_{0}\left( k\right) &=&-2t_{1}\cos k_{x}\beta _{x}\alpha _{0}-2t_{1}\cos
k_{y}\beta _{y}\alpha _{z}+m_{0}\beta _{z}\alpha _{0}  \notag \\
&&+4t_{2}\sin k_{x}\sin k_{y}\beta _{z}\alpha _{0}+\beta _{z}\left( %
\bm{\mathit{h}}\cdot \bm{\mathit{\alpha}}\right)
\end{eqnarray}%
under the basis $\hat{C}_{k}=\left( \hat{c}_{\text{A},\uparrow ,k},\hat{c}_{%
\text{B},\uparrow ,k},\hat{c}_{\text{A},\downarrow ,k},\hat{c}_{\text{B}%
,\downarrow ,k}\right) ^{T}$. Here, $t_{1}$ and $t_{2}$ denote the tunneling
strength between nearest-neighbor and next-nearest-neighbor sites. $%
\bm{\mathit{\alpha}}$ and $\bm{\mathit{\beta}}$ represent Pauli matrices
acting on spin ($\uparrow $, $\downarrow $) and sublattice ($\text{A}$, $%
\text{B}$) degrees of freedom, respectively. $\alpha _{0}$ denotes a
two-by-two identity matrix. $m_{0}$ represents the staggered on-site
potential at two sublattices. $\bm{\mathit{h}}$ is an in-plane Zeeman energy
that are opposite at two sublattice sites. Hereafter, without loss of
generality, we set $\bm{\mathit{h}}=(h_{0},0,0)$ if not specified.

\begin{figure}[tbp]
\centering\includegraphics[width=0.46\textwidth]{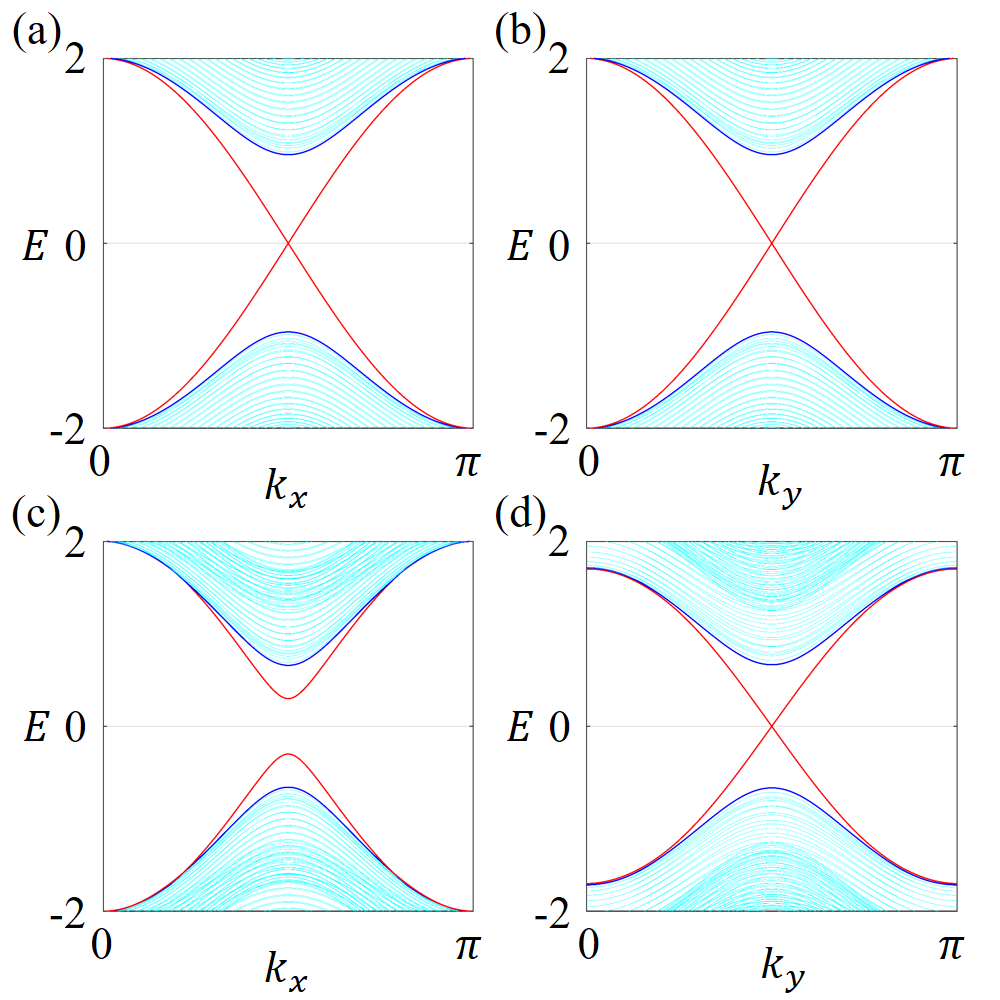}
\caption{(a) and (c) Energy spectra under periodic boundary condition in the
$x$ direction but open boundary condition in the $y$ direction. (b) and (d)
Energy spectra under periodic boundary condition in the $y$ direction but
open boundary condition in the $x$ direction. In (a) and (b) $h_{0}=0.0$,
and in (c) and (d) $h_{0}=0.3$. Common parameters are $t_{1}=1$, $t_{2}=0.25$%
, $m_{0}=0.05$.}
\label{edgeTI}
\end{figure}

When $h_{0}=0$, the system preserves time reversal symmetry, and describes a
quantum spin-Hall insulator in the band inverted region $%
t_{2}^{2}>m_{0}^{2}/16$. This nontrivial phase is characterized by $Z_{2}$
topological invariant and hosts helical gapless edge states, as shown in
Fig. \ref{edgeTI} (a) and (b). After turning on OIPZE $\bm{\mathit{h}}$, the
time-reversal symmetry is broken for the system. However, the edge states
remain gapless along $y$ while are gapped along $x$ as illustrated in Fig. %
\ref{edgeTI} (c) and (d). This phenomenon originates from that OIPZE
commutes with the nearest-neighbor term along $y$, but anti-commutes with
the nearest-neighbor term along $x$. It would be more explicit from the
effective low-energy edge theory. See details in Sec. \ref{S5}.

Consider the on-site attractive interaction for fermions defined by $\hat{H}%
_{U}=-U\sum_{i}n_{i,\uparrow }n_{i,\downarrow }$, and now the Hamiltonian
for this correlated system becomes
\begin{equation}
\hat{H}=\sum_{k}\hat{C}_{k}^{\dagger }H_{0}\left( k\right) \hat{C}_{k}+\hat{H%
}_{U}-\mu \sum_{i}n_{i},  \label{eq1}
\end{equation}%
where $\mu $ denotes the chemical potential, and $n_{i}=n_{i,\uparrow
}+n_{i,\downarrow }$. In the following, we study the ground phases for this
system through the mean-field theory and the effective low-energy theory.

\section{Phase diagram}

\label{S3}As the attractive interaction increases, it is expected that
fermions would be paired into a superfluid phase if the superfluid gap
function exceeds the energy gap of single-particle Hamiltonian $H_{0}$.
Therefore we introduce the mean-field ansatz for the superfluid order
parameter as $\Delta _{s}=-U\left\langle \hat{c}_{i,\uparrow }^{\dagger }%
\hat{c}_{i,\downarrow }^{\dagger }\right\rangle $. Through Fourier
transformation, the kernel of the Hamiltonian in momentum space is written
as
\begin{eqnarray}
H\left( k\right) &=&\kappa _{k}\beta _{z}\alpha _{0}\gamma _{z}-2t_{1}\cos
k_{x}\beta _{x}\alpha _{0}\gamma _{z}-2t_{1}\cos k_{y}\beta _{y}\alpha
_{z}\gamma _{0}  \notag \\
&&+\triangle _{s}\beta _{0}\alpha _{y}\gamma _{y}-\mu _{0}\beta _{0}\alpha
_{0}\gamma _{z}+h_{0}\beta _{z}\alpha _{x}\gamma _{z}
\end{eqnarray}%
under the basis $\hat{\Psi}_{k}=\left( \hat{C}_{k},\hat{C}_{-k}^{\dagger
}\right) ^{T}$, where $\kappa _{k}=4t_{2}\sin k_{x}\sin k_{y}-m_{0}$, $%
\bm{\mathit{\gamma}}$ represents Pauli matrices acting on particle-hole
space, $\beta _{0}$ and $\gamma _{0}$ are two-by-two identity matrices. The
free energy for the ground state at zero-temperature for the system is given
by $F_{u}=-2\sum_{k}(E_{k,+}+E_{k,-})+2N_{u}\triangle _{s}^{2}/U$, where $%
E_{k,\pm }=\sqrt{\left( \epsilon _{k}\pm h_{0}\right) ^{2}+\left( 2t_{1}\cos
kx\right) ^{2}}$ with $\epsilon _{k}=\sqrt{\kappa _{k}^{2}+\left( 2t_{x}\cos
ky\right) ^{2}}$, and $N_{u}$ is the number of unit cells. By minimizing the
ground state energy $F_{u}$ with respect to the superfluid order $\Delta _{s}
$, we obtain the self-consistent equation as%
\begin{equation}
\frac{U}{2N_{s}}\sum_{k}\left( \frac{h_{0}/\zeta _{k}+1}{E_{k,+}}-\frac{%
h_{0}/\zeta _{k}-1}{E_{k,-}}\right) =1  \label{eqs}
\end{equation}%
with $\zeta _{k}=\sqrt{\kappa _{k}^{2}+\left( 2t_{x}\cos ky\right)
^{2}+\triangle _{s}^{2}}$. Here we consider the case with $\mu =0$ for
simplicity but without loss of generality. Through numeric calculations
based on Eq. (\ref{eqs}), we get the phase diagram for the system, as shown
in Fig. \ref{phases} (b). It shows that as attractive interaction increases
from zero, the system remains an insulator below the critical interaction
values. When the attractive interaction exceeds critical values, the system
enters the superfluid phase.

\begin{figure}[tbp]
\centering\includegraphics[width=0.46\textwidth]{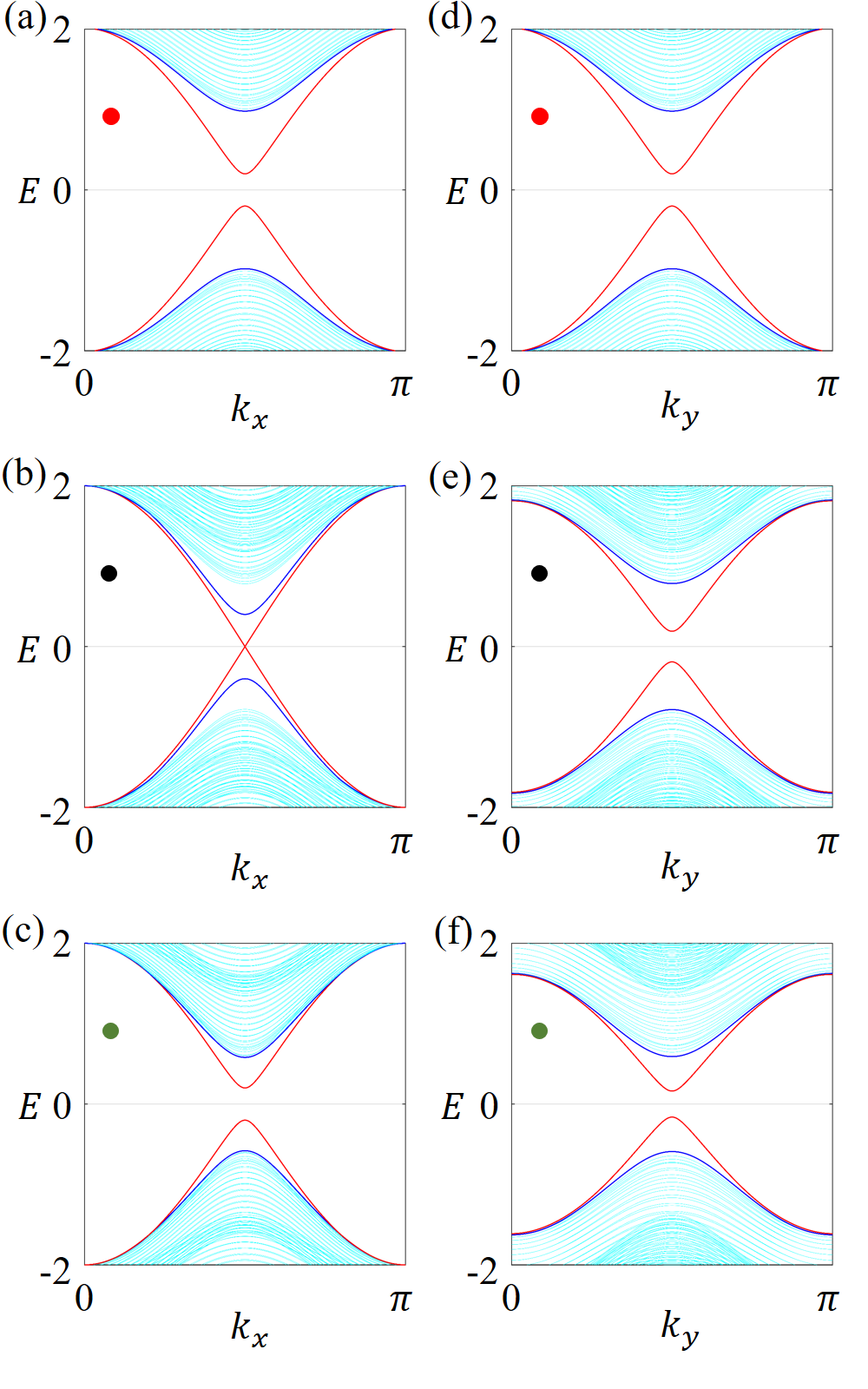}
\caption{(a)-(c) Energy spectra for a strip under periodic boundary
conditions along $x$ but open boundary conditions along $y$. (d)-(e) Energy
spectra for a strip under periodic boundary conditions along $y$ but open
boundary conditions along $x$. OIPZE $h_{0}$ is $0$ in (a) and (d), $0.2$ in
(b) and (e), and $0.4$ in (c) and (f). Common parameters are set to be $%
t_{1}=1$, $t_{2}=0.25$, $m_{0}=0.05$, $\triangle _{s}=0.2$, $\protect\mu %
_{0}=0$. These chosen parameters in sub-figures also have been indicated by
colored dots in Fig. \protect\ref{phases}(b).}
\label{edge2}
\end{figure}

We next investigate topological properties of superfluid phases. OIPZE plays
different roles on the helical edge states (HES) of quantum spin Hall
insulators. We would like to see the response of gapless HES to the
combination of OIPZE and superfluid order. We first plot energy spectra for
a stripe under periodic boundary conditions along $x$ ($y$) but open along $%
y $ ($x$) with increasing OIPZE and fixed superfluid order parameter $%
\triangle _{s}=0.2$, as shown in Fig. \ref{edge2}. It showcases that when $%
h_{0}=0$, the edge states along both $x$ and $y$ are gaped by the superfluid
order. As illustrated in Figs. \ref{edge2}(a)-(c), the gap for edge states
along $x$ decreases gradually as OIPZE increases, closes at $h_{0}=\triangle
_{s}$ and reopens when $h_{0}>\triangle _{s}$. However, the gap for edge
states along $y$ remains gaped in this process, as shown in Figs. \ref{edge2}%
(d)-(f). This implies that a topological phase transition from a trivial to
a nontrivial phase may occur for edges states along $x$ while no topological
phase transition occurs for edge states along $y$.

To demonstrate that the system is in a topological phase when $%
h_{0}>\triangle _{s}$, we compute energies of the superfluid on a square
lattice under open boundary conditions along $x$ and $y$, as presented in
the inset of Fig. \ref{den} (a). It shows that there are four Majorana
zero-energy states in the energy gap. Their density distributions, as shown
in Fig. \ref{den}(a), indicate that they are localized at four corners of
the sample, which implies that this superfluid phase is a second-order
topological superfluid (SSF) hosting Majorana corner states. Therefore, to
summarize, the SF phase consists of a SSF and a normal SF, as shown in the
phase diagram in Fig. \ref{phases}(b).

\section{Topological invariants}

\label{S4} We utilize Majorana edge polarizations to characterize
topological properties for this second-order topological superfluid \cite%
{yjwu2020}. We first construct Wilson loop $\mathcal{W}=F_{k_{x}+2\pi
}...F_{k_{x}+\Delta k}F_{k_{x}}$, where $F_{k_{x}}$ is a $4N_{y}$-by-$4N_{y}$
matrix with the matrix element $\left( F_{k_{x}}\right) _{mn}=\left\langle
u_{k_{x}+\Delta k_{x}}^{m}|u_{k_{x}}^{n}\right\rangle $, where $\left\vert
u_{k_{x}}^{n}\right\rangle $ is $n$th occupied Bloch wave function with
eigen-energy $E_{n}$, $N_{y}$ is the site number along $y$, and the momentum
interval $\Delta k_{x}=2\pi /N_{x}$ with $N_{x}$ the number of unit cells.
The Wannier Hamiltonian for $\mathcal{W}$ is then defined by
\begin{equation}
\mathcal{H=}\frac{1}{2\pi i}\ln \mathcal{W}.
\end{equation}%
The eigenvalues of $\mathcal{H}$ are Wannier values $\nu^{x} _{j}$
corresponding to eigenfunctions $\left\vert \nu _{k_{x}}\right\rangle $. By
using Bloch wave functions, Wannier values and corresponding eigenfunctions,
we define Majorana edge polarization along $y$ as $p_{y}^{edge,x}=%
\sum_{i_{y}=1}^{N_{y}/2}p_{i_{y}}^{x}$, where
\begin{equation}
p_{i_{y}}^{x}=\frac{1}{N_{x}}\sum_{k_{x},m,j}\nu^{x} _{j}\left(
_{m}\left\langle \nu _{k_{x}}^{j}\right\vert _{i_{y}}\left\langle
u_{k_{x}}^{m}|u_{k_{x}}^{m}\right\rangle _{i_{y}}\left\vert \nu
_{k_{x}}^{j}\right\rangle _{m}\right) .
\end{equation}%
Majorana edge polarization $p_{x}^{edge,y}$ along $x$ takes similar
formulation as $p_{y}^{edge,x}$.

\begin{figure}[tbp]
\centering\includegraphics[width=0.485\textwidth]{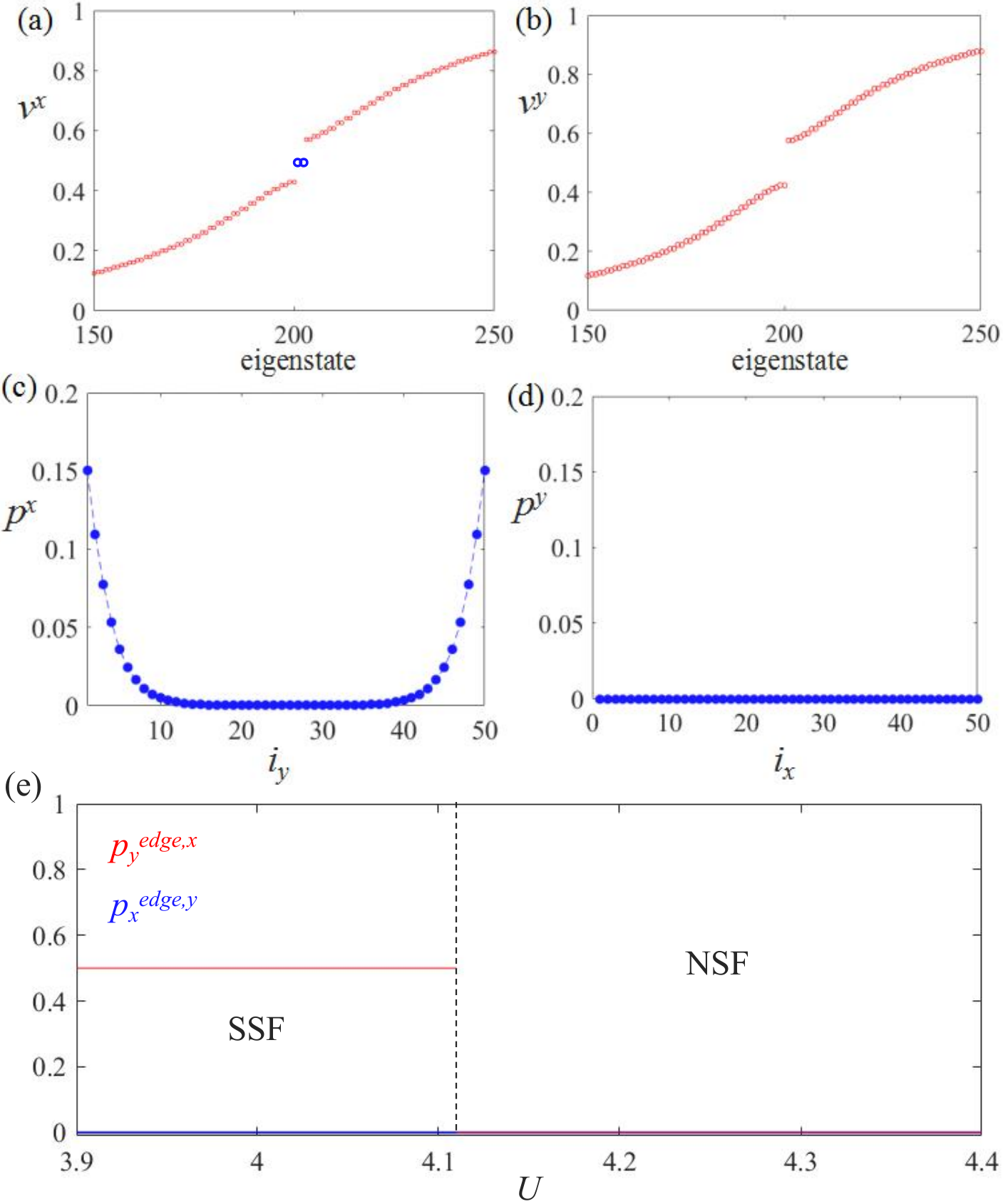}
\caption{(a) Wannier values $\protect\nu_{x}$ for a strip under periodic
boundary conditions along $x$ but open boundary conditions along $y$. (b)
Wannier values $\protect\nu_{y}$ for a strip under periodic boundary
conditions along $y$ but open boundary conditions along $x$. (c) and (d)
Majorana edge polarizations at each lattice site along $y$ and $x$. In
(a)-(d), $\triangle _{s}=0.2$. (e) Majorana edge polarizations versus
interaction $U$ in SSF and NSF phases. Common parameters are $t_{1}=1$, $%
t_{2}=0.25$, $h_{0}=0.4$ $m_{0}=0.05$ and $\protect\mu _{0}=0$.}
\label{polar}
\end{figure}

In SSF phase, two quantized Wannier values $\nu^{x}=0.5$ emerge in the gap
of Wannier band $\nu^{x}$ while not in the Wannier band $\nu^{y}$, as shown
in Fig. \ref{polar}(a) and (b). It shows that edge states along $x$ exhibit
non-trivial topological properties. We next further compute Majorana edge
polarizations on each site, as ploted in Fig. \ref{polar}(c) and (d).
Obviously, nontrivial Majorana edge polarization exists for edge states
along $x$ while not for edge states along $y$. Through numeric calculations,
we obtain the topological invariants $(p_{y}^{edge,x},p_{x}^{edge,y}=(1/2,0)$
for SSF phase. While for NSF phase, the topological invariant becomes $(0,0)$%
, as shown in Fig. \ref{polar}(e). From the perspective of Majorana edge
polarizations, the emergence of Majorana corner states could be simply
understood as follows. Half quantized Majorana edge polarization along $x$
implies that edge states along $x$ are topologically nontrivial. Yet, zero
Majorana edge polarization along $y$ indicates that edge states along $y$
are topologically trivial. When trivial and nontrivial edge states encounter
at a corner, Majorana zero-energy corner state naturally arises from the
index theorem. Finally, with the consideration of the topological phase
transition in the superfluid phase, we plot the global phase diagram in Fig. %
\ref{phases}(b).

\begin{figure}[tbp]
\centering\includegraphics[width=0.48\textwidth]{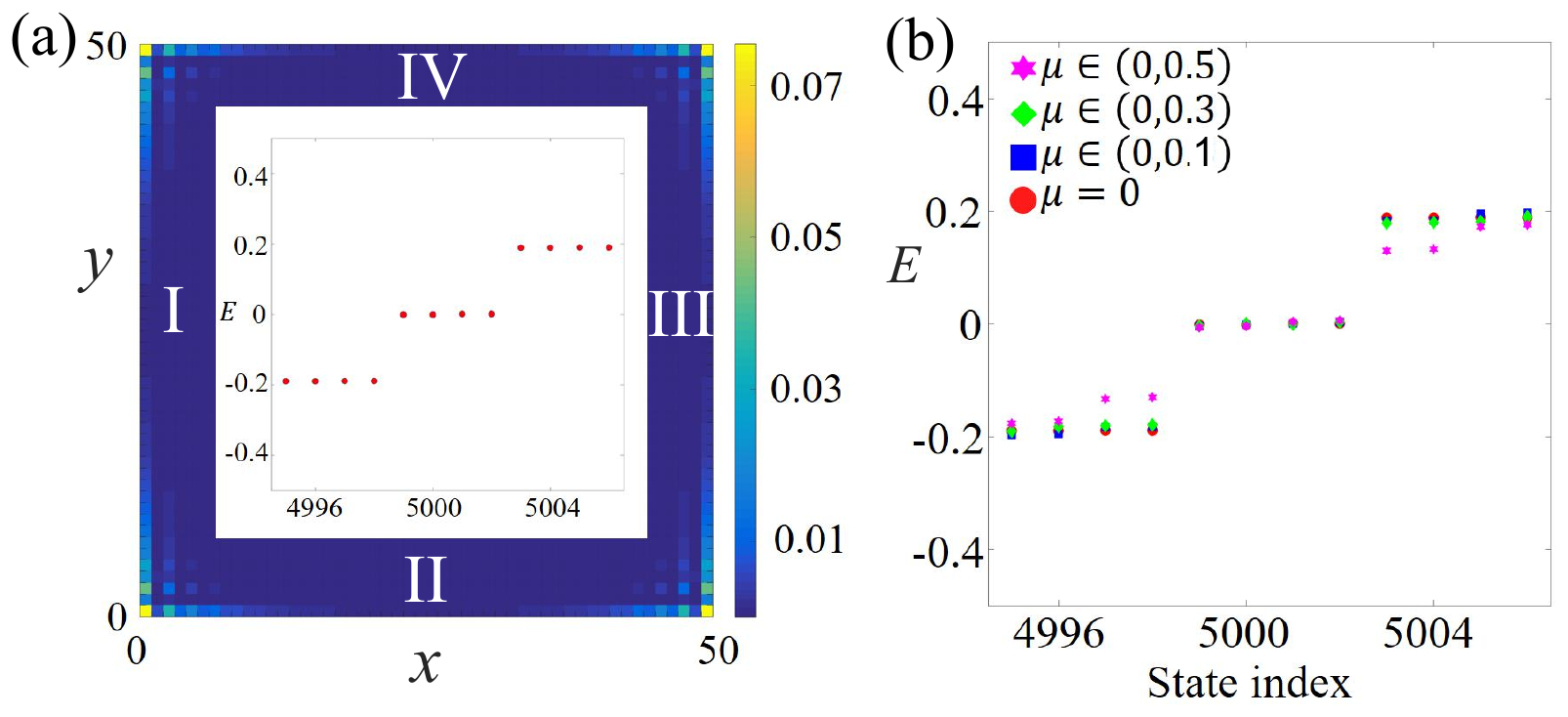}
\caption{(a) Density distributions of Majorana zero-energy states on a $50*50
$ square lattice. The inset shows energies of the square sample. Parameters
are chosen as $h_{0}=0.4$, $\triangle_{s}=0.2$, $\protect\mu _{0}=0$. (b)
Energies of SSF in the presence of random fluctuations on chemical
potentials. Common parameters are set to be $t_{0}=1$, $t_{1}=0.25$, $%
m_{0}=0.05$.}
\label{den}
\end{figure}

To show the robustness of Majorana corner states, we impose random
fluctuations on chemical potentials. The energies of a square sample in the
presence of random fluctuations with different amplitudes have been shown in %
\ref{den}(b). It showcases Majorana corner states are robust against
perturbations as long as the edge gap remains open.

\section{Effective low-energy edge theory}

\label{S5}

In this section, we will explore the emergence of Majorana corner states
form low-energy edge theory. The low-energy Hamiltonian in the continuum
limit around $K=\left( \pi /2,\pi /2\right) $ up to the second order is
written as%
\begin{eqnarray}
H_{\mathrm{LE}}(k) &=&\left( \varepsilon
-2t_{2}k_{x}^{2}-2t_{2}k_{y}^{2}\right) \beta _{z}\alpha _{0}\gamma
_{z}+2t_{1}k_{x}\beta _{x}\alpha _{0}\gamma _{z}  \notag \\
&&+2t_{1}k_{y}\beta _{y}\alpha _{z}\gamma _{0}+\Delta _{s}\beta _{0}\alpha
_{y}\gamma _{y}+h_{0}\beta _{z}\alpha _{x}\gamma _{z}
\end{eqnarray}%
with $\varepsilon =4t_{2}-m_{0}$. As illustrated in Fig. \ref{den}(a), the
four edges of a square lattice are labeled by $\mathrm{I}$, $\mathrm{II}$, $%
\mathrm{III}$ and $\mathrm{IV}$. For edge $\mathrm{I}$, we replace the
momentum operator $k_{x}$ by $-i\partial _{x}$. The Hamiltonian then becomes
$H_{\mathrm{LE}}(k)=H_{M,\mathrm{I}}+H_{P,\mathrm{I}}$, where the main term $%
H_{M,\mathrm{I}}$ and perturbation term $H_{P,\mathrm{I}}$ are given by%
\begin{eqnarray}
H_{M,\mathrm{I}}\left( -i\partial _{x},k_{y}\right)  &=&\left( \varepsilon
+2t_{2}\partial _{x}^{2}\right) \beta _{z}\alpha _{0}\gamma
_{z}-2it_{1}\partial _{x}\beta _{x}\alpha _{0}\gamma _{z},  \notag \\
H_{P,\mathrm{I}}\left( -i\partial _{x},k_{y}\right)  &=&-t^{\prime
}k_{y}^{2}\beta _{z}\alpha _{0}\gamma _{z}+2t_{y}k_{y}\beta _{y}\alpha
_{z}\gamma _{0}+\Delta _{s}\beta _{0}\alpha _{y}\gamma _{y}  \notag \\
&&+h_{0}\beta _{z}\alpha _{x}\gamma _{z}.
\end{eqnarray}%
Here we have assumed that the pairing order compared to the energy gap is
relatively small .

In the following, we first solve the main part $H_{M,\mathrm{I}}$, and then
derive the effective low-energy edge Hamiltonian for edge $\mathrm{I}$. We assume $H_{M,\mathrm{I}}$ has zero energy solutions $%
\psi _{a}$ localized at edge $\mathrm{I}$. Since $\left\{ H_{M,\mathrm{I}%
},\beta _{y}\alpha _{z}\gamma _{z}\right\} =0$, $\beta _{y}\alpha _{z}\gamma
_{z}\psi _{m}$ are also eigen-states for $H_{M,\mathrm{I}}$. Therefore we
choose eigenvectors $\chi _{m}$ with $\beta _{y}\alpha _{z}\gamma _{z}\chi
_{m}=-$ $\chi _{m}$, where
\begin{eqnarray}
\chi _{1} &=&\left\vert \beta _{y}=-1\right\rangle \left\vert \alpha
_{z}=+1\right\rangle \left\vert \gamma _{z}=+1\right\rangle ,  \notag \\
\chi _{2} &=&\left\vert \beta _{y}=+1\right\rangle \left\vert \alpha
_{z}=-1\right\rangle \left\vert \gamma _{z}=+1\right\rangle ,  \notag \\
\chi _{3} &=&\left\vert \beta _{y}=+1\right\rangle \left\vert \alpha
_{z}=+1\right\rangle \left\vert \gamma _{z}=-1\right\rangle ,  \notag \\
\chi _{4} &=&\left\vert \beta _{y}=-1\right\rangle \left\vert \alpha
_{z}=-1\right\rangle \left\vert \gamma _{z}=-1\right\rangle .
\end{eqnarray}%
Four zero-energy states $\psi _{m=1,2,3,4}$ localize at edge \textrm{I} in
this basis with $\ \psi _{m}=\mathcal{N}\sin \vartheta e^{-\frac{t_{1}}{%
2t_{2}}x}\chi _{m}$, where $\mathcal{N}$ is a normalization constant, and\ $%
\vartheta =\sqrt{\varepsilon /\left( 2t_{2}\right) -t_{1}^{2}/\left(
4t_{2}^{2}\right) }$. In the basis $\psi _{m}$, the effective low-energy
edge Hamiltonian for the perturbation term $H_{P,\mathrm{I}}$ then becomes%
\begin{equation}
H_{\mathrm{Edge,I}}=it_{y}\alpha _{z}\gamma _{0}\partial _{y}+\Delta
_{s}\alpha _{y}\gamma _{y}+h_{0}\alpha _{x}\gamma _{z}.
\end{equation}%
The low-energy effective Hamiltonian at edges \textrm{II}, \textrm{III},
\textrm{IV} takes similar formulations as
\begin{eqnarray}
H_{\mathrm{Edge,II}} &=&-it_{1}\alpha _{z}\gamma _{0}\partial _{x}+\Delta
_{s}\alpha _{y}\gamma _{y},  \notag \\
H_{\mathrm{Edge,III}} &=&-it_{1}\alpha _{z}\gamma _{0}\partial _{y}+\Delta
_{s}\alpha _{y}\gamma _{y}+h_{0}\alpha _{x}\gamma _{z},  \notag \\
H_{\mathrm{Edge,IV}} &=&it_{1}\alpha _{z}\gamma _{0}\partial _{x}+\Delta
_{s}\alpha _{y}\gamma _{y}.
\end{eqnarray}%
To summarize, we have%
\begin{equation}
H_{\mathrm{Edge}}=it_{1}\alpha _{z}\gamma _{0}\partial _{x}+\Delta
_{s}\alpha _{y}\gamma _{y}+h\left( m\right) \alpha _{x}\gamma _{z},
\label{10}
\end{equation}%
with $h\left( m\right) =h_{0}, 0, h_{0}, 0$ at edges $m=\mathrm{I}, \mathrm{%
II}, \mathrm{III}, \mathrm{IV}$, respectively. To be more explicit, we
rewrite $H_{\mathrm{Edge}}$ as%
\begin{equation}
H_{\mathrm{Edge}}^{\prime }=-it_{1}\alpha _{z}\gamma _{0}\partial
_{x}+\Delta _{s}\alpha_{x}\tau _{z}+h\left( m\right) \alpha _{x}\gamma _{0}
\end{equation}%
in the space $\eta _{1}=\left\vert \alpha _{z}=+1\right\rangle \left\vert
\gamma _{x}=+1\right\rangle $, $\eta _{2}=\left\vert \alpha
_{z}=-1\right\rangle \left\vert \gamma _{x}=-1\right\rangle $, $\eta
_{3}=\left\vert \alpha _{z}=+1\right\rangle \left\vert \gamma
_{x}=-1\right\rangle $, $\eta _{4}=\left\vert \alpha _{z}=-1\right\rangle
\left\vert \gamma _{x}=+1\right\rangle $. Now $H^{\prime}_{\mathrm{Edge}}$
is composed of two $2$-by-$2$ decoupled blocks, one block with Dirac mass $%
h\left( m\right) +\Delta _{s}$, and the other with $h\left( m\right) -\Delta
_{s}$. When Dirac masses on adjacent edges have opposite signs, i.e., $%
\left( h_{0}+\Delta _{s}\right) .\Delta _{s}<0$ or $\left( h_{0}-\Delta
_{s}\right) .\left( -\Delta _{s}\right) <0$, Majorana zero-energy state
naturally arises at the corner.

If a general OIPZE $\bm{\mathit{h}}=(h_{x},h_{y},0)$ is applied, the term $%
h\left( m\right) \alpha _{x}\gamma _{z}$ in Eq.(\ref{10}) would be replaced
by another term $h\left( m\right) \tilde{\alpha}_{\theta}\tau _{0}$, where $%
\left\{ \alpha _{z},\tilde{\alpha}_{\theta }\right\} =0$ and $%
h\left(m\right)=h, 0, h, 0$ with $h=\left \vert{\bm{\mathit{h}}}\right\vert $
at edges $\mathrm{I}-\mathrm{IV}$, respectively. Majorana zero states
naturally emerge when $\left( h+\Delta _{s}\right).\Delta _{s}<0$ or $\left(
h-\Delta _{s}\right) .\left( -\Delta _{s}\right) <0$. Hence, with the
assistance of $s$-wave superfluid order and appropriate OIPZE at two
sublattice sites, Dirac mass of edge states on adjacent edges possesses opposite signs
which naturally induces Majorana zero-energy states at corners.

\section{Discussion and conclusion}

\label{S6}

The ultracold atom system provides a highly controllable platform with
various degrees of freedom. The quantum spin Hall insulator in Eq.(\ref{eq1}%
) could be implemented by a spin-dependent optical lattice \cite%
{Hou2013,YJWu2016}, and the OIPZE can be reached by another addressing
laser. The attractive interaction can be tuned by Feshbach resonance
technique \cite{CChin2010,TKohler2006}. Using the generalized Bragg
spectroscopy and the exapansion of the atomic cloud \cite%
{Goldman2013,Goldman2012}, one may detect the induced Majorana corner states.

In summary, in contrast to previous schemes, we propose an OIPZE can drive a
second-order topological superfluid phase in an attractive quantum spin Hall
insulator. Here, the cooperation of OIPZE and the s-wave superfluid order
leads to the opposite signs of Dirac mass at adjacent edges. Majorana zero states
naturally emerge localized at corners. They are robust against weak
perturbations as long as the edge energy gap remains open. This topological
phase can be characterized by the Majorana edge polarizations. Our work
provides a new route towards Majorana corner states through the cooperation
of the OIPZE and the s-wave pairing.

\begin{acknowledgments}
This work is supported by the Scientific Research Program Funded by the
Natural Science Basic Research Plan in the Shaanxi Province of China
(Programs No. 2021JM-421 and No. 2019JM-001), the NSFC under the Grant No.
11504285, the Scientific Research Program Funded by Shaanxi Provincial
Education Department under the grant No. 18JK0397, and the Young Talent fund
of the University Association for Science and Technology in Shaanxi, China
(Program No. 20170608).
\end{acknowledgments}

\newpage \clearpage\onecolumngrid\appendix

\end{document}